\documentstyle[aps,pra,epsfig,twocolumn]{revtex}

\begin{document}
\draft

\title{Entanglement sharing among qudits}

\author{Kenneth A.~Dennison and William K.~Wootters}

\address{Department of Physics, Williams College, Williamstown, 
MA 01267, USA}

\maketitle

\begin{abstract}
Consider a system consisting of $n$ $d$-dimensional quantum 
particles (qudits), and suppose that we want to optimize
the entanglement between each pair.  
One can ask
the following basic question regarding the sharing of 
entanglement: what is the largest 
possible value $E_{max}(n,d)$ of the
{\em minimum} entanglement between any two 
particles in the system?  (Here we take the entanglement
of formation as our measure of entanglement.)  
For $n=3$ and $d=2$, that is, for a system of three qubits,
the answer is known: $E_{max}(3,2) = 0.550$.  
In this paper
we consider first a system of $d$ qudits and show that
$E_{max}(d,d) \geq 1$.  We then consider a system of
three particles, with three different values of $d$.
Our results for the three-particle case 
suggest that as the dimension $d$ increases,
the particles can share a greater 
fraction of their entanglement
capacity.  
\end{abstract}


\narrowtext

Quantum entanglement, as exhibited, for example, in the singlet state 
$(1/\sqrt{2})(|01\rangle - |10\rangle)$ of a pair of qubits,
has been the object of much study
in recent years because of its connection with quantum 
communication and quantum computation \cite{Nielsen}.  
Though entanglement is a kind of correlation, it is
known to be fundamentally
different from ordinary classical correlation.
One of the characteristic differences 
is this: whereas
arbitrarily many classical systems can be perfectly 
correlated with each other---the temperature fluctuations
in ten different cities could in principle be exactly 
parallel---any entanglement that may exist
between two quantum particles seems to limit
the degree to which either of
the particles can be entangled with anything 
else \cite{Bruss,Coffman}.
For example, if two qubits $A$ and $B$ are in the singlet state,
then neither of them can have any entanglement
with a third qubit $C$, simply because such entanglement
would require the pair $AB$ to be in a {\em mixed} state,
whereas the singlet state is pure.  
Coffman {\em et al.} \cite{Coffman} have
generalized this example (still considering only
qubits) by allowing 
$A$ and $B$ to be
only partially entangled, in which case one finds an inequality
expressing a trade-off between the $AB$-entanglement and 
the $AC$ or $BC$-entanglement.   

As this sort of limitation may be a fundamental
property of entanglement, one would like to express
it more generally.  
In particular, one would like to capture quantitatively the limitation
on the sharing of entanglement among arbitrarily many particles
of arbitrary dimension.  
The following problem offers one approach to such 
a quantitative expression.  Consider a system of
$n$ $d$-dimensional quantum particles (qudits), and suppose 
that one wants each particle to be highly entangled with each
of the other particles. We expect that there will have to
be compromises, since increasing the entanglement
of any given pair will probably work against the entanglements
of other pairs.  It makes sense, then, to ask how large
one can make the {\em minimum} pairwise 
entanglement, the minimum being taken over
all pairs \cite{Dur}.
In this paper we address this problem, taking as
our measure of entanglement the entanglement
of formation \cite{BDSW,BBPSSW}, which for a pair of 
qudits ranges from 
zero to $\log_2 d$.  
For a collection of $n$ qudits, let
us call the maximum possible 
value of the minimum pairwise entanglement
$E_{max}(n,d)$.  This function, if it can be found, will
give us a specific quantitative bound
on the degree to which entanglement can be shared 
among a number of particles.  We focus in this paper
on two special 
cases: $n=d$ and $n=3$.  As we will see, our results
for $n=3$,
combined with earlier work on the problem, suggest
that in a well defined sense 
the limitation on entanglement sharing
becomes less restrictive with increasing values of
the dimension $d$.  

Before reviewing what is currently known 
about $E_{max}(n,d)$, let us recall 
the definition of entanglement
of formation.  
For a pure state $|\Phi\rangle$ of a 
bipartite quantum system, the entanglement $E(\Phi)$
is defined \cite{BBPS} as 
\begin{equation}
E(\Phi) = -\sum_i r_i \log_2 r_i,
\end{equation}
where the $r_i$'s are the eigenvalues of the density
matrix of either subsystem.  
(For a pure bipartite state the density 
matrices of both subsystems necessarily
have the same eigenvalues.)
A mixed state $\rho$ can always be written in
many different ways as a probabilistic mixture of distinct
but not necessarily orthogonal pure states:
\begin{equation}
\rho = \sum_j p_j |\Phi_j\rangle\langle\Phi_j|.
\end{equation}
The entanglement of formation of $\rho$ is 
defined \cite{BDSW,BBPSSW}
as the average entanglement of the pure states 
of the decomposition, minimized over all possible decompositions:
\begin{equation}
E_f(\rho) = \inf \sum_j p_j E(\Phi_j).
\end{equation}
As we have mentioned above, the entanglement of
formation between a pair of qudits ranges from 
zero to $\log_2 d$.  Let us refer to
the maximum value $\log_2 d$ as the {\em entanglement
capacity} of a pair of qudits.

For a pair of qubits, there is an explicit formula
for the entanglement of formation of an arbitrary
mixed state \cite{W}.  It is given in terms of another measure
of entanglement called the concurrence \cite{W,HW}, which
at this point has a standard definition 
only for qubits.\footnote{The concurrence of a pure
state of two qubits is simply $2\sqrt{\det \rho_A}$,
where $\rho_A$ is the reduced density matrix of
one of the qubits.  The concurrence of a mixed state $\rho$
of two qubits is given by $\max \{0,\lambda_1 - \lambda_2
-\lambda_3 -\lambda_4\}$, where $\lambda_1\geq
\lambda_2\geq\lambda_3\geq\lambda_4$ are the square
roots of the eigenvalues of $\rho(\sigma_y\otimes\sigma_y)
\rho^*(\sigma_y\otimes\sigma_y)$, $\rho^*$ being the
complex conjugate of $\rho$ in the standard basis and
$\sigma_y$ being the usual Pauli matrix
\cite{W,HW}.  We recall these formulas here for the 
sake of completeness
but will not need them in the present paper.}  
In terms of the concurrence $C$, which ranges from
zero to one, the
entanglement of formation of a pair of qubits is
$E_f(\rho) = {\cal E}(C(\rho))$, where the function
${\cal E}$ is defined by
\begin{equation}
{\cal E}(C) = h\left[\frac{1}{2}\Big(1+\sqrt{1-C^2}\Big)\right], \label{E(C)}
\end{equation}
$h$ being the binary entropy function
$h(x) = -x\log_2 x - (1-x)\log_2 (1-x)$.  
Note that ${\cal E}(C)$ is a monotonically increasing
function, with ${\cal E}(0)=0$ and ${\cal E}(1)=1$.
We will not be focusing particularly on qubits in
this paper,
but Eq.~(\ref{E(C)}) will be useful both for summarizing
previous work on the problem and, in a different context,
for presenting our own results.

We now list the results that have
been obtained so far regarding $E_{max}(n,d)$.
\begin{enumerate}
\item $E_{max}(2,d) = \log_2 d$.  This equation simply says that
if there are only two particles, they can saturate their 
entanglement capacity;
they do not have to share the entanglement with other
particles.
\item $E_{max}(3,2) = {\cal E}(2/3) = 
0.550$.  D\"ur {\em et al.}~\cite{Dur} obtained
this result by proving that the optimal pairwise entanglement
for a system of three qubits is achieved in the state
$(1/\sqrt{3})(|100\rangle + |010\rangle + |001\rangle)$.
\item $E_{max}(n,2) \geq  {\cal E}(2/n)$.   
Koashi {\em et al.} \cite{Koashi}
showed that for a system of $n$ qubits, if the state 
is such that the density 
matrix of each pair of particles is the same, then the 
maximum pairwise concurrence is $2/n$.  It is conceivable
(though it seems unlikely) that by removing the 
symmetry constraint
one might be able to achieve a greater pairwise entanglement;
therefore we write this result as an inequality rather than
an equality.  
\end{enumerate}
In this paper we add two new items to the above list: 
(i) For $n=d$, that is, for a system of $d$ qudits,
we find for each value of $d$ 
a specific state in which each pair of particles
has exactly 1 ``ebit'' of entanglement between them.
(For $d=2$, our state reduces to the singlet state
of a pair of qubits.)
This will show that $E_{max}(d,d)$ is at least 1 for all
values of $d$.  (ii) For $n=3$, that is,
for a system of three particles, we add to the known
result for qubits ($d=2$) and to our own result for
qutrits ($d=3$ in item (i)) a third example with $d=7$.
Our results for the three-particle case suggest that
as $d$ increases, the particles 
can share not just more
entanglement, but a greater {\em fraction} of their
entanglement capacity. 

\bigskip

\noindent {\bf A system of $d$ qudits}

\medskip

\noindent Before writing down our special state of $d$ qudits
with arbitrary $d$, we illustrate our 
construction in
the special case of three qutrits.  
Let the particles be called $A$, $B$ and $C$, and 
let the indices $i$, $j$ and $k$ label the elements of orthogonal
bases for the three particles,
each index taking the values 0, 1 and 2. 
Our special state for this system is
\begin{equation}
|\xi\rangle = \frac{1}{\sqrt{6}}
\sum_{i,j,k} \epsilon_{ijk}|ijk\rangle,
\end{equation}
where $\epsilon_{ijk}$ is antisymmetric under
interchange of any two indices and $\epsilon_{012} = 1$.
This is the singlet state of three qutrits with respect to 
the group $SU(3)$; {\em i.e.}, it is the 
unique three-qutrit state (up to an overall phase factor) 
that is
invariant under arbitrary transformations of the form
$U\otimes U\otimes U$ where $U \in SU(3)$.
The density matrix $|\xi\rangle\langle\xi|$ is 
symmetric under interchange
of any two particles, so that
each pair of particles is equally entangled.  To find
the pairwise entanglement, we write down the 
reduced density matrix of any pair; for definiteness
we choose the first two particles, $A$ and $B$:
\begin{equation}
\rho^{AB}_{ij,i'j'} = \frac{1}{6}\sum_k \epsilon_{ijk}
\epsilon_{i'j'k} = \frac{1}{6}(\delta_{ii'}\delta_{jj'}
-\delta_{ij'}\delta_{ji'}),
\end{equation}
$\delta$ being the Kronecker delta.
Alternatively, we can write $\rho^{AB}$ without indices
as
\begin{equation}
\rho^{AB} = \frac{1}{6}(I - F),
\end{equation}
where $I$ is the identity operator
and $F$ is the operator that interchanges particles $A$ and $B$:
$F = \sum_{ij} |ij\rangle\langle ji|$.   
The two-qutrit state $\rho^{AB}$ is an example of a
Werner state, that is, a state that is invariant under
all transformations of the form $U \otimes U$ where $U$
is unitary.  Werner states can be defined for any $d\times d$
system, and one can show \cite{VW} that in any dimension the Werner
states are precisely those states that can be written as
$\rho = aI + bF$, $a$ and $b$ being real numbers and $F$
being defined as above.   Vollbrecht
and Werner \cite{VW} have shown that the entanglement of 
formation of any Werner state is given by 
$E_f(\rho) = {\cal E}(c(\rho))$, where $c(\rho)
= -\hbox{Tr}\,\rho F$ and  
${\cal E}$ is the function defined in Eq.~(\ref{E(C)}).  
(When $c(\rho)$ is non-negative,
it plays the role of a concurrence for Werner states.)
In our case, $c(\rho^{AB}) = 1$, so that the entanglement
is $E_f(\rho^{AB}) = {\cal E}(1) = 1$.  Thus each pair of 
qutrits has exactly one ebit of entanglement.  This value is,
by the way,
the maximum possible entanglement of any Werner state.

It is a simple matter to generalize the above construction
to a system of $d$ qudits.  In that case, we have $d$
indices $i_1, i_2, \ldots, i_d$, each taking values from
0 to $d-1$.  Our special state for this system is the
$SU(d)$ singlet state\footnote{This
state has been used recently by Hillery and Bu\v{z}ek in a 
scheme designed to probe a quantum gate that realizes
an unknown unitary transformation \cite{Hillery}.}
\begin{equation}
|\xi\rangle = \frac{1}{\sqrt{d!}}\sum_{i_1\ldots i_d}
\epsilon_{i_1\ldots i_d}|i_1\ldots i_d\rangle,
\end{equation}
where $\epsilon_{i_1\ldots i_d}$ is completely
antisymmetric and $\epsilon_{0,1,\ldots,d-1} = 1$.
One can show directly that the reduced density matrix
of each pair of particles is again a Werner state:
\begin{equation}
\rho^{AB} = \frac{1}{d(d-1)} (I - F),
\end{equation}
and that the entanglement of formation of this 
state is one ebit.  We thus conclude that
$E_{max}(d,d) \geq 1$.  We write an inequality 
here simply
because our state $|\xi\rangle$ may not
optimize the pairwise entanglement; one
might be able to do better.  However, having
put some effort into looking for better states
with $d=3$, we regard it as likely that
our state is optimal in that case.

\bigskip

\noindent {\bf A three-particle system}

\medskip

It is interesting to compare our result for three
qutrits with the previously studied example of 
three qubits \cite{Dur}.  For a triple of qubits
$E_{max}$ is 0.550, and we have just
seen that for a triple of qutrits, $E_{max}$ is at 
least 1.  However, a straightforward comparison
of these numbers is not particularly illuminating,
because qubits and qutrits have different entanglement
capacities. 
We can perhaps make a fairer comparison by
considering the ratio of $E_{max}$ to 
the relevant entanglement
capacity.  For qubits, this ratio is $0.550/\log_2 2 = 0.550$,
whereas for qutrits it is $1/\log_2 3 = 0.631$.  Thus by this
measure, 
qutrits are 
better able to share entanglement than qubits: they
can share a greater fraction of their entanglement
capacity.  
It is interesting to ask whether this trend will continue for 
larger values of $d$.  That is, will $E_{max}(3,d)/\log_2 d$ continue
to increase with increasing $d$?  

To address this question, we consider one further
case with three particles, namely, the case $d=7$.  We
choose the value 7 because it allows us to construct
a reasonably simple and symmetric state that exhibits
large pairwise entanglement.  
In fact we consider a one-parameter
{\em family} of states, having the
following form:
\begin{equation}
|\zeta\rangle = \frac{1}{\sqrt{7}}  \label{Z}
\sum_{j=0}^6 \Big(a|j,j,j\rangle
+ b\sum_{k\in Q} |j+k,j+2k,j+4k\rangle \Big).    
\end{equation}
Here $Q$ is the set $\{1,2,4\}$, and all the
arithmetic shown in the ket labels is  
mod 7, the basis
states of each particle being labeled by the 
integers $0,\ldots,6$.  We take
$a$ and $b$ to be real and positive, with
$a^2 + 3b^2 =1$ to ensure normalization.  
Thus the state $|\zeta\rangle$ is completely 
specified once the value of $a$ is given.  

We have chosen $Q$ to consist of the
{\em quadratic residues} mod 7, that is, the 
elements of $\{1,2,3,4,5,6\}$ that can be 
written as $x^2$ mod 7 for some integer $x$.
The properties of quadratic 
residues \cite{Mac} tend
to minimize the overlap, in each particle's state
space, between terms in Eq.~(\ref{Z}) with
different values of $j$.  (For this it is important
that 7 is a prime of the form $4N-1$ with 
integral $N$.)

A particular symmetry of $|\zeta\rangle$
shows immediately that the pairs $AB$, $BC$, and
$CA$ are all equally entangled.  If we define a new
summation index $k'$ by $k = 2k'$ mod 7, then 
Eq.~(\ref{Z}) becomes
\begin{equation}
 |\zeta\rangle = \frac{1}{\sqrt{7}}  \label{prime}
\sum_{j=0}^6 \Big(a|j,j,j\rangle
+ b\sum_{k'\in Q} |j+2k',j+4k',j+k'\rangle \Big),    
\end{equation}
where we have used the invariance of $Q$ 
under multiplication by 2 mod 7.  But Eq.~(\ref{prime})
differs from Eq.~(\ref{Z}) in that the ket labels
have been cyclically permuted.  Thus $|\zeta\rangle$
is invariant under a cyclic permutation of the particles,
and it follows that each pair is equally entangled.  

To write down the density matrix of one of the pairs, say $BC$,
it is helpful to re-express Eq.~(\ref{Z}) in yet another
form, changing the index $j$ in the $jk$-sum to $j'=j+k$
and then relabeling $j'$ as $j$:
\begin{equation}
|\zeta\rangle = \frac{1}{\sqrt{7}}  
\sum_{j=0}^6 \Big(a|j,j,j\rangle
+ b\sum_{k\in Q} |j,j+k,j+3k\rangle \Big). 
\end{equation}
The density matrix of $BC$ is the trace of
$|\zeta\rangle\langle\zeta|$ over particle $A$,
which we can write as 
\begin{equation}
\rho_{BC} = \frac{1}{7}\sum_{j=0}^6 |s_j\rangle\langle s_j|,
\end{equation}
where $|s_j\rangle$, defined by 
\begin{equation}
|s_j\rangle = a|j,j\rangle + b\sum_{k\in Q} |j+k,j+3k\rangle,
\end{equation}
is the state of $BC$ associated with the state $|j\rangle$
of $A$.  

In order to find the entanglement of formation $E(\rho_{BC})$,
we need to consider pure-state decompositions of $\rho_{BC}$
and find their average entanglements.  
Now, any pure state $|\beta\rangle$ 
in such a decomposition must be a linear
combination of the seven orthogonal 
states $|s_j\rangle$ that make up
$\rho_{BC}$; that is, it must lie in the seven-dimensional
subspace ${\mathcal H}$ spanned by $\{|s_j\rangle \}$:
\begin{equation}
|\beta\rangle = \sum_j \beta_j |s_j\rangle,
\end{equation}
where $\sum_j |\beta_j|^2 = 1$.  The problem of finding
$E(\rho_{BC})$ is simplified by two facts: (i) $E(\rho_{BC})$
cannot be smaller than the smallest entanglement of
any $|\beta\rangle \in {\mathcal H}$; that is, $E(\rho_{BC}) \geq \min_{\beta}E(\beta)$.
(ii) Given any state $|\beta\rangle \in {\mathcal H}$, 
one can generate
an entire decomposition of $\rho_{BC}$ in which {\em every} element has the
same entanglement as $|\beta\rangle$.  (We prove this assertion in the 
following paragraph.)  Together, these two facts imply that the 
entanglement of formation of $\rho_{BC}$ 
is {\em equal} to $\min_{\beta}E(\beta)$.
Thus it is sufficient to find a single minimally entangled pure state in
the subspace ${\mathcal H}$ occupied by $\rho_{BC}$.  

To generate a decomposition of $\rho_{BC}$ from a given state
$|\beta\rangle \in {\mathcal H}$, 
we apply a set of local unitary transformations
to $|\beta\rangle$; such transformations are guaranteed not to
change the entanglement.  We start by defining two
basic single-particle transformations $S$ and $T$:
\begin{equation}
S|j\rangle = \omega |j\rangle; \hspace{0.4cm} T|j\rangle = |j+1\rangle,
\end{equation}
where $\omega = e^{2\pi i/7}$ and, as always, the addition in the ket label
is mod 7.  In terms of these basic operations, we define a pair of two-particle
transformations $U$ and $V$:
\begin{equation}
U = S^5 \otimes S^3; \hspace{0.4cm} V = T \otimes T.
\end{equation}
One can show that $U|s_j\rangle = \omega^j |s_j\rangle$ and
$V|s_j\rangle = |s_{j+1}\rangle$, from which it follows that
\begin{eqnarray}
&\frac{1}{49}\sum_{m=0}^6 \sum_{p=0}^6 
V^p U^m |\beta\rangle\langle\beta| U^{-m}V^{-p} \\ 
&=\frac{1}{49}\sum_{j,j'}\sum_{m,p} \omega^{(j-j')m} \beta_j \beta_{j'}^*
|s_{j+p}\rangle \langle s_{j'+p}|  \\
&= \frac{1}{7} \sum_j |\beta_j|^2 
\sum_p |s_{j+p}\rangle\langle s_{j+p}| = \rho_{BC}.
\end{eqnarray}
We have thus produced the desired decomposition of $\rho_{BC}$.

It remains, then, to find the smallest possible value of $E(\beta)$.  For the
special case where $a = b = 1/2$, 
each state $|s_j\rangle$ has exactly two bits of 
entanglement, but it happens that
certain linear combinations of the
states $|s_j\rangle$ have slightly smaller entanglement.
Using numerical minimization, we find
that for this case, $\min E(\beta) = 1.9933$.  

Of course we are free to choose the value of $a$ as we please, and it turns out 
that we maximize the entanglement of formation
by choosing $a = 0.461$,
in which case $b = 0.512$.  For this value of $a$, we find 
numerically that the 
minimum $E(\beta)$ is 1.9944, obtained both for the simple case 
$|\beta\rangle = |s_j\rangle$
and for certain non-trivial linear combinations.  [One such combination
has coefficients
$\beta_j$ equal to $(0.120, 0.197, 0.689, 0.259, -0.468, -0.275, -0.332)$, and the others we have found are all related to this one
by permutations and phase changes.]  We conclude, then, that for the state $|\zeta\rangle$
with $a = 0.461$, the entanglement of formation between each
pair of particles is 1.9944, and therefore $E_{max}(3,7) \geq 1.9944$.    

This result gives us another data point as we consider the dependence
of the ratio $E_{max}(3,d)/\log_2 d$ on the dimension $d$.  The 
following table summarizes what we know so far about
the case $n=3$.  (For $d=3$ and $d=7$, the values given are lower bounds.)  

\medskip

\begin{center}
\begin{tabular}{|c|c|c|}
\hline  
$d$  &  $E_{max}^{(bound)}(3,d)$  &  $E_{max}^{(bound)}(3,d)/\log_2 d$ \\
\hline  
2  &  0.550  &  0.550  \\
3  &  1.000  &  0.631  \\
7  &  1.994  &  0.710  \\
\hline
\end{tabular}
\end{center}

\medskip

In the limit as $d$ goes to infinity, we wonder what value,
if any, the ratio $E_{max}(3,d)/\log_2 d$ approaches.  
It is conceivable
that the limit is 1, but it is equally conceivable that it is
some smaller constant.  Either answer would be interesting.  If
the limit of $E_{max}(n,d)/\log_2 d$ is 1 for all values of $n$, 
then one could reasonably say that 
entanglement can be shared freely in an infinite dimensional 
state space.  

We note that although in this
paper we have 
focused on the entanglement of formation, 
there exist other sound
measures of entanglement, and
it is surely
a good idea, in trying to quantify the restrictions 
on the sharing
of entanglement, 
to keep in mind alternatives
such as the 
relative entropy of entanglement\cite{rel}
and the generalized
concurrence of Rungta {\em et al.} \cite{Rungta}.
At the present stage of investigation, it is not
clear which measure or measures will yield
the most elegant quantitative expressions 
of the limitations
on entanglement sharing.

\end{document}